\documentclass[twocolumn,showpacs,preprintnumbers,amsmath,amssymb,superscriptaddress]{revtex4}
\usepackage{appendix}
\usepackage{amsmath,amssymb,amsfonts}
\usepackage{graphics}
\usepackage{dcolumn} % Align table columns on decimal point
\usepackage{bm,graphicx,color}
\usepackage{float}

\usepackage{braket}
\begin{document}

\title{
Dy adatom on MgO(001) substrate: DFT+U(HIA) study
}
%LDA+Exact Diagonalization Calculations of Co Impurity in Cu Bulk}

\author{Alexander B. Shick}
\affiliation{Institute of Physics, Czech Academy of Sciences, Na Slovance 2, 182 21 Prague,
Czech Republic.}
\affiliation{
Department of Molecular Chemistry and Materials Science, Weizmann Institute of Science, Rehovoth 76100, Israel.}

\author{Eduard Belsch}
\affiliation{Institute of Physics, Czech Academy of Sciences, Na Slovance 2, 182 21 Prague,
Czech Republic.}
\affiliation{Institute of Theoretical Physics, University of Hamburg, 20355 Hamburg, Germany}

\author{Alexander I. Lichtenstein}
\affiliation{Institute of Theoretical Physics, University of Hamburg, 20355 Hamburg, Germany}
\affiliation{European X-Ray Free-Electron Laser Facility, Holzkoppel 4, 22869 Schenefeld, Germany}

\begin{abstract}
The electronic structure and magnetism of  individual Dy atom adsorbed on the MgO(001) substrate is
 investigated using the combination of the density functional theory with the Hubbard-I approximation to the Anderson impurity model (DFT+U(HIA)).
 The divalent  Dy$^{2+}$  adatom in $f^{10}$ configuration is found. The calculated  x-ray absorption (XAS) and magnetic circular dichroism (XMCD) spectra are compared to the experimental data.
 Quantum tunneling between degenerate $\ket{J=8.0, J_z=  \pm 4.0 }$ states leads to formation of   $\ket{J=8.0, J_z= 0.0 }$ ground state with an in-plane orientation of the magnetic moment. It explains 
 absence of remanent magnetization in MgO adatom on the top of Mg(001) substrate. 
 Our studies can provide a viable route for further investigation and prediction of the rare-earth single atom  magnets.
\end{abstract}
\date{\today}

%\pacs{71.20,71.27+a,75.40.Cx} 

\maketitle
%\section
%{\em Introduction.}
Lantanide atom adsorption on suitable surfaces is a viable pathway for creating atomic scale magnetic memories~\cite{Donati2021L} and quantum logic 
devices~\cite{Thiele2014}. Dysprosium (Dy) exibits a large magnetic anisotropy and can be protected against quantum tunneling in a uniaxial crystal field~\cite{Singha2021}. It has been used for molecular magnets with record-high blocking temperature~\cite{goodwin2017}, and the surface adsorbed single atom magnets with the long magnetization lifetime~\cite{Baltic2016}.

Recently, it was shown experimentally~\cite{Donati2021} that the electronic properties of Dy adatoms on MgO thin films
grown on the top of metal Ag(001) substrate change with the thickness of supporting MgO layer. 
X-ray absorption spectroscopy (XAS), and magnetic circular dichroism (XMCD) at 2.5 K reveal a predominance of the bulklike 4$f^9$ Dy for the Dy@MgO/Ag(001) with the MgO layer thickness less 
than 5 monolayers. By an increase of the MgO layer thickness, Dy atoms  acquire the 4$f^{10}$
configuration. They display the butterfly-type magnetic hysteresis loop, indicating quantum tunneling of the magnetization (QTM). 

Despite the relatively simple coordination of the atom support structure, it remains challenging to predict
theoretically an influence of the substrate and adsorption geometry on 
the Dy 4$f$-shell charge and magnetic configurations. Theoretical calculations
often require a prior knowledge of the experimental data~\cite{Donati2021}.  
The density functional theory (DFT) is used to obtain the optimized 
adsorption geometry. The XAS spectra are then fitted making use of MultiX multiplet calculations~\cite{Uldry2012}
together with a point charge
model with the positions and values of the Born charges
deduced from DFT. 
%Importantly, these calculations require a prior knowledge
%of the experimental XAS spectra.

In this work, we present an alternative  theoretical approach, 
based on the combination of relativistic DFT with the multiorbital impurity Hamiltonian,
and apply it to investigate the electronic and magnetic character of Dy adatom at MgO(001). 
Our calculations suggest that the multiconfigurational aspect of the Dy 4$f$-shell 
together with a correct atomic limit need to be taken into account in order to reproduce 
the  magnetic and spectroscopic properties of  Dy@MgO.  
%The earlier version of the
%method was used to treat the magnetic properties of Dy adatom deposited on the Cu and Ir
%supported graphene~\cite{shick2020}.

%\section
%{\em Method.}
The DFT+U correlated electronic structure theory in a rotationally invariant, 
full potential implementation~\cite{shick99,shick01}, minimizes the total energy functional
\begin{equation}
E^{tot}(\rho,\hat{n}) =  E^{DFT}(\rho) + E^{ee}(\hat{n}) - E^{dc}(\hat{n}) \; , 
\label{totale}
\end{equation}
where, $E^{DFT}(\rho)$ is usual density functional
of the total electron and spin densities, 
$\rho({\bf r})$, including SOC. $E^{ee}$ is an electron-electron interaction energy
and $E^{dc}$ is a ``double-counting'' term which accounts approximately
for an electron-electron interaction
energy already included in $E^{DFT}$. Both are functions of
the local orbital occupation matrix 
$\hat{n} = n_{\gamma_1 \gamma_2}$
in the subspace of the $f$ spin-orbitals 
$\{\phi_{\gamma} =  \phi_{m \sigma} \}$.

Minimization of the DFT+U total energy functional Eq.~\ref{totale}
leads to the solution of the generalized  Kohn--Sham-Dirac
equations, 
\begin{equation}
\bigl[ -\nabla^{2} + {V}_{\rm DFT}(\mathbf{r}) + (V_{U} - V_{dc}) + \xi ({\bf l} \cdot
{\bf s}) \bigr]  \Phi_{\bf k}({\bf r}) = \epsilon_{\bf k} \Phi_{\bf
k}({\bf r}),
\label{eq:kohn_sham}
\end{equation}
where, ${V}_{U}$ is an effective DFT+U
potential, and $V_{dc}$ is the spherically-symmetric DFT+U  double-counting term~\cite{AZA1991,solovyev1994}
%either in the so-called around-mean-field  
%(AMF) 
%$V_{dc} =  (U/2 \; {n_f } + \frac{2l}{2(2l+1)} \; (U-J) \; %{n_f})$
% form, or the fully localized limit (FLL)
%$V_{dc} =  (U-J)/2 \;  ({n_f } - 1)$
%\cite{solovyev1994}. 
The self-consistent solition of in Eq.(\ref{eq:kohn_sham})
generates not only the ground state energy and charge/spin densities, 
but also effective one-electron states and energies.
%(the band structure).
The basic difference of DFT+U calculations from DFT is its explicit dependence
on the on-site spin- and orbitally resolved  $n_{\gamma_1 \gamma_2}$ occupation matrices .

%In DFT+U, the electron interaction effects beyond DFT are accounted as
%the on-site atomic-like correlations. 
The fundamental limitation of DFT+U calculations 
is that they rely on a single Slater determinant approximation for the $f$-manifold. 
However, as pointed out in Ref.~\cite{shick2001,Dorado2013}, it makes the DFT+U results extremely sensitive to the initial conditions, which leads to numerious metastable solutions. 

In order
to avoid convergence to a metastable state, various strategies have been proposed. The occupation matrix control (OMC) has recently been exploited by Krack~\cite{Krack2015}
for the two $f$-electrons, however the identified ground state does not agree with earlier DFT+U results of Dorado {\em et al.}~\cite{Dorado2010}. Alternatively, the so-called $U$-ramping 
method relies on a gradual increase of the Coulomb-$U$ parameter of DFT+U. While this approach has had some success, it has been shown to give higher energies than the OMC method~\cite{Meredig2010}.

Recently, we proposed the extention of DFT+U~\cite{SFP2021} making use of a combination
of DFT with the exact diagonalization of the Anderson impurity model~\cite{Hewson}.
The complete seven-orbital 4$f$ shell model includes the full spherically symmetric
Coulomb interaction, the spin-orbit coupling, and the crystal
field. The corresponding Hamiltonian can be written as,
\begin{align}
\label{eq:hamilt}
\hat{H}_\text{imp}  = & 
%\sum_{\substack {k m m' \\ \sigma \sigma'}}
% [\epsilon^{k}]_{m m'}^{\sigma \; \; \sigma'} b^{\dagger}_{km\sigma}b_{km'\sigma'} +
 \sum_{m\sigma} \epsilon_f f^{\dagger}_{m \sigma}f_{m \sigma} \\
& + \sum_{mm'\sigma\sigma'} \bigl[\xi {\bf l}\cdot{\bf s}
  + \hat{\Delta}_{\rm CF} + \frac{\Delta_{\rm EX}}{2} \hat{\sigma}_z\bigr]_{m m'}^{\sigma \; \; \sigma'}
  f_{m \sigma}^{\dagger}f_{m' \sigma'}
\nonumber \\
& + \frac{1}{2} \sum_{\substack {m m' m''\\  m''' \sigma \sigma'}}
  U_{m m' m'' m'''} f^{\dagger}_{m\sigma} f^{\dagger}_{m' \sigma'}
  f_{m'''\sigma'} f_{m'' \sigma},
\nonumber
\end{align}
where $f^{\dagger}_{m \sigma}$ creates a 4$f$ electron.
The $\xi$ parameter specifies the SOC strength, and
is taken from DFT
calculations in a standard way~\cite{MPK1980}, making use of the radial solutions of the Kohn-
Sham-Dirac scalar-relativistic equations~(\ref{eq:kohn_sham}), 
%the relativistic mass M = m + (El − V (r))/2c2 at an appropriate energy El, 
and the radial derivative of spherically-
symmetric part of the DFT potential.
$\Delta_{\rm CF}$ is the 
crystal-field potential, and $\Delta_{\rm EX}$ is the exchange field strength. 
The parameter $\epsilon_f$ ($= -\mu$, the  chemical
potential) defines the number of $f$-electrons.  The last term describes the
Coulomb interaction in the $f$-shell. Actual choice of these parameters will be discussed later.

This model  assumes the weakness of the hybridization between the  localized $f$-electrons and the itinerant $s$, $p$, and $d$-states described in DFT. Thus, the quantum impurity Anderson model~\cite{Hewson} is reduced to the atomic limit, and corresponds to the Hubbard-I approximation (HIA). 

%We approximate the $s$, $p$, and $d$ shells in DFT.

The Lanczos method~\cite{Kolorenc2012} is employed to find
the lowest-lying eigenstates of the many-body Hamiltonian $H_\text{imp}$  and to calculate the selfenergy matrix $[\Sigma (z)]_{\gamma, \gamma'}$ 
in the subspace of the $f$ spin-orbitals 
$\{\phi_{\gamma} =  \phi_{m \sigma} \}$
at low temperature ($k_{\rm B}T=\beta^{-1} = 2$ meV). 
Once the selfenergy is found, the local Green's function $G(z)$ for
the electrons in the 4$f$ manifold reads,
\begin{eqnarray}
\label{eq:gf}
G(z) =  \Big( [{G(z)}_{\rm DFT}]^{-1} + \Delta \epsilon - \Sigma(z) \Big)^{-1} \, , 
\end{eqnarray}
where  ${G}_{\rm DFT}(z)$ is the ``non-interacting" DFT Green's function, and $\Delta \epsilon$ is chosen so as to ensure
that $n_f \; = \; - \pi^{-1} {\rm Im} \; {\rm Tr}\int_{-\infty}^{E_{\rm{F}}} {\rm d} z  [ G(z)]$ is
equal to the number of 4f electrons derived from Eq.~\eqref{eq:kohn_sham}.
Then, with the aid of the local Green's
function $G(z)$, we evaluate
the occupation matrix
$n_{\gamma_1 \gamma_2} = - \pi^{-1} \mathop{\rm Im}
\int_{-\infty}^{E_{\rm{F}}} {\rm d} z \, [G(z)]_{\gamma_1 \gamma_2}$.

This matrix $n_{\gamma_1 \gamma_2}$ is used to construct an effective DFT+U
potential ${V}_{U}$ in Eq.(\ref{eq:kohn_sham}). 
Note that the DFT  potential
${V}_{\rm DFT}$ in Eq.(\ref{eq:kohn_sham}) acting on the $f$-states is corrected to exclude  the non-spherical double-counting with $V_U$~\cite{Kristanovski2018}. 
The equations Eq.(\ref{eq:kohn_sham}) are iteratively solved until self-consistency over
the charge density is reached. The new DFT Green's function  ${G}_{\rm DFT}$ and  the 
new value of the 5$f$-shell occupation are  obtained from the
solutions of Eq.~(\ref{eq:kohn_sham}).  
The next iteration is started by solving~Eq.~(\ref{eq:hamilt}) 
with the updated value of $\epsilon_f = -\mu$ in Eq.~(\ref{eq:hamilt}),
which is determined by the condition $\mu = V_{dc}$~\cite{SFP2021}.

%This is an essential condition. The double-counting term $V_{dc} $ accounts 
%approximately for the electron-electron interaction energy $E^{ee}_{\rm DFT}$ already
%included in the DFT.  Thus it can be written as 
%a derivative of this mean energy contribution with respect to the $f$-shell occupation $n_f$,
%$V_{dc}=\partial E^{ee}_{\rm DFT}/\partial n_{f}$. Indeed, it represents
%a value of the chemical potential $\mu$ that controls
%the number of $f$ electrons.
 The loop procedure  is repeated until the
convergence of  the 4$f$-manifold occupation $n_f$ is better than
0.02. After the self-consistent solution of DFT+U(HIA) is obtained, 
the mean-field total energy $E_{Tot} = E_{\rm DFT} + \Delta E^{ee} $ is calculated as a sum of DFT  total energy $E_{\rm DFT}$, 
and the energy correction $\Delta E^{ee} \; = \; E^{ee} \; - \; E_{dc}$. Importantly, this solution is unique as it stems from the many-body
ground state of ~Eq.~(\ref{eq:hamilt}) with the exact atomic limit.
%(the difference between the electron-electron interaction energy
%$E^{ee}$  and double-counting energy $E_{dc}$ already included in $E_{\rm DFT}$).

\begin{figure}[h]
\centerline{\includegraphics[angle=0,width=1.0\columnwidth,clip]{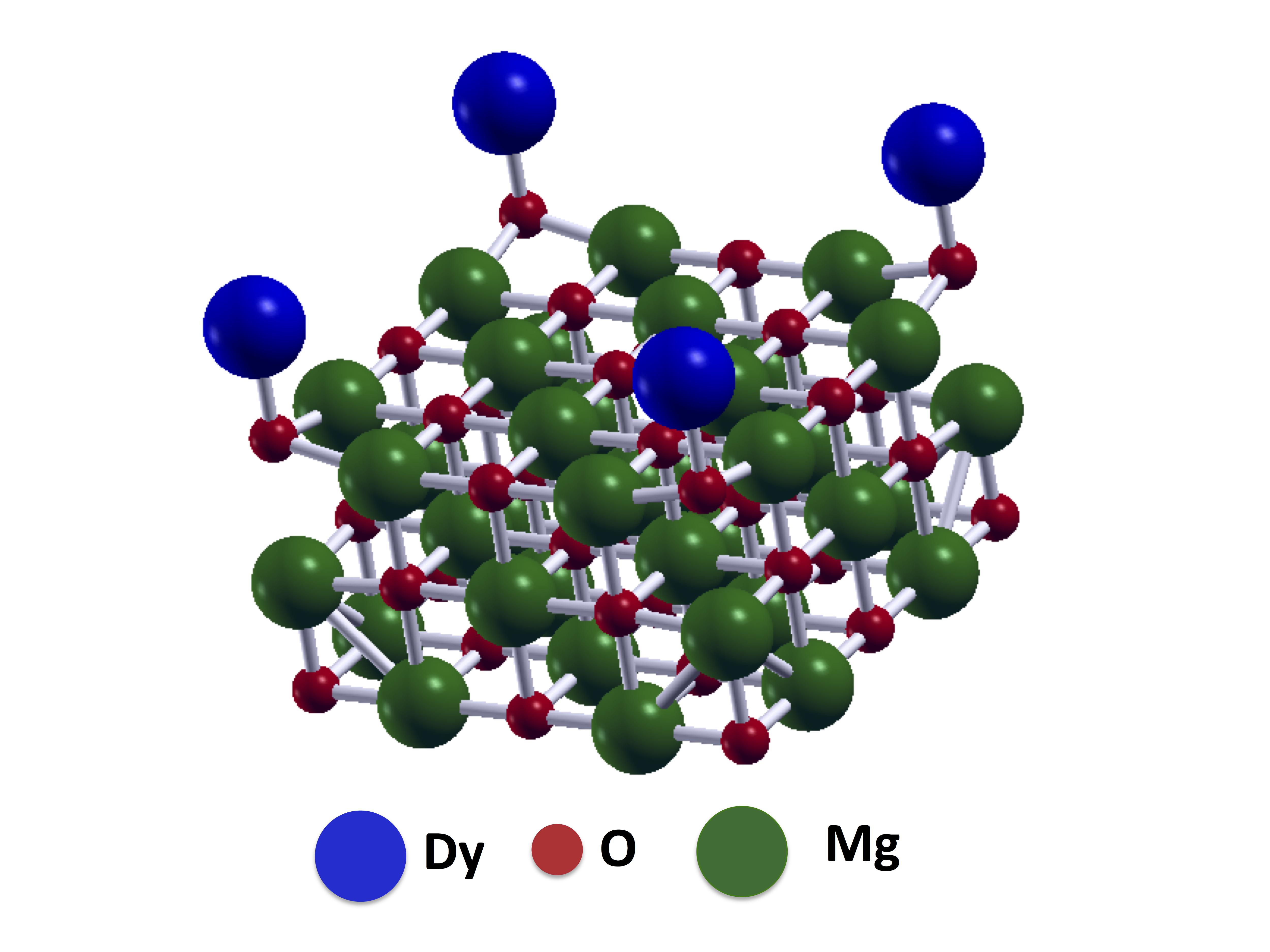}}
\caption{Supercell model for rare-earth impurity on MgO(001).
Dy atoms are shown in blue, O atoms are in red, and Mg atoms in green.}
 \label{fig:1}
\end{figure}

%\section
%{\em Results and discussion}
We make use of the $2\times2\times1$ lateral supercell ($a=4.21$ \AA) of 3 ML of MgO to which the rare-earth Dy adatom is added on the oxyden site. In order to obtain the supercell geometry, we performed the standard DFT 
(with the exchange-correlation functional of Perdew, Burke and Ernzerhof~\cite{PBE}) Vienna ab initio simulation package (VASP~\cite{VASP}) calculations together with the projector augmented-wave method (PAW~\cite{PBE}).
Moreover, assuming that localised 4$f$ electrons have rather small impact on the geometry,
we used the rare-earth Lu adatom instead of Dy, and treated 14 closed 4$f$-shell electrons of Lu as valence. The system was relaxed until the forces on the Lu adatom and on top-most 2 ML of MgO are  0.001 eV/{\AA}. The calculated 2.1 {\AA} Lu-O bond length to the underneath oxygen is in a good quantitative agreement with the DFT+U results of Ref.~\cite{Donati2021} for Dy-O bond length.  Calculated adsorption geometry is shown in the Fig.~\ref{fig:1}. 

The structural information obtained from the VASP
simulations was used as an input for further DFT+U(HIA)
electronic structure calculations that employ the relativistic version of the full-potential linearized augmented plane-wave method (FP-LAPW)~\cite{FLAPW}. In the FP-LAPW  the SOC is included in a
self-consistent second-variational procedure~\cite{shick1997}. This two-step approach synergetically combines the speed and ef-
ficiency of the highly optimized VASP package with the
state-of-the-art accuracy of the FP-LAPW method

%In the DFT+U(HIA) FP-LAPW calculations, 25 special k-points in the two-dimensional  Brillouin zone
%were used, with Gaussian smearing for k-points weighting.
%The ``muffin-tin'' radii of $R_{MT} \; = \; 2.70 \; a.u.$ for Dy,
%$R_{MT}=1.20$ a.u. for O,  $R_{MT}=2.00$ a.u. for Mg  were used.
%The LAPW basis cut-off is defined by the condition
%$R^{Dy}_{MT} \times K_{max} \; = \; 8.10$
%(where $K_{max}$ is the cut-off for LAPW basis set).  
The Slater integrals  $F_0=7.00$~eV, and $F_2=9.77$ eV, $F_4=6.53$ eV, and $F_6=$4.83 eV were chosen
to parametrize the Coulomb interaction term in Eq.~(\ref{eq:hamilt}), and to construct the DFT+U potential ${V}_{U}$ in the Eq.~(\ref{eq:kohn_sham}).
They corresponds to the values for Coulomb~$U=7.00$~eV and exchange~$J = 0.82$ eV.
The above choice of the Slater integrals is justified~\cite{shick2020} 
by agreement between the density of states (DOS) calculated with DFT+U(HIA) and the experimental
valence band photoemission for the bulk Dy. 

%\subsection{Evaluation of $\Delta_{EX}$}
\begin{figure}[!htbp]
\centerline{\hspace{-1cm} \includegraphics[angle=0,width=1.0\columnwidth,clip]{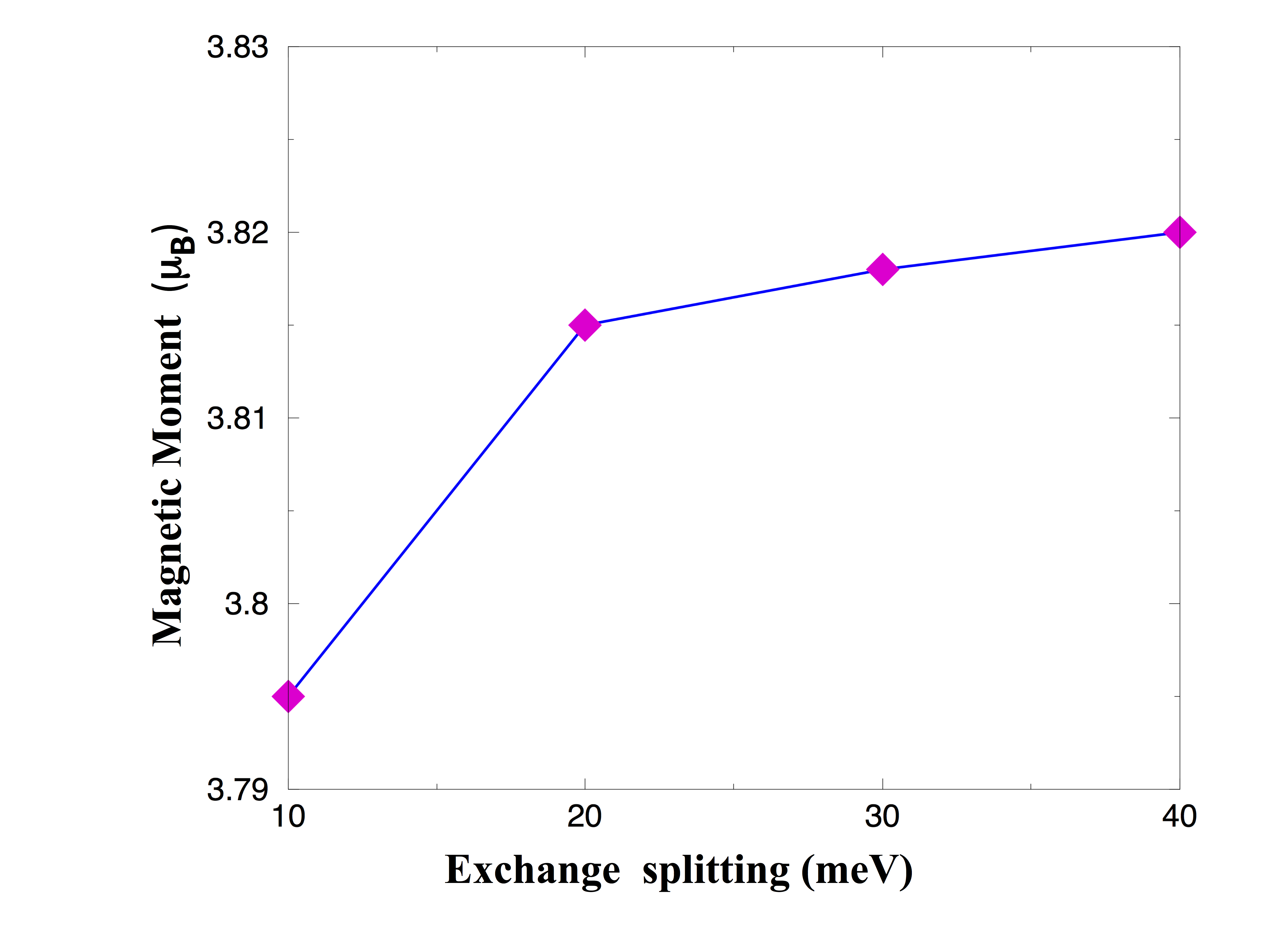}}
\centerline{\includegraphics[angle=0,width=1.0\columnwidth,clip]{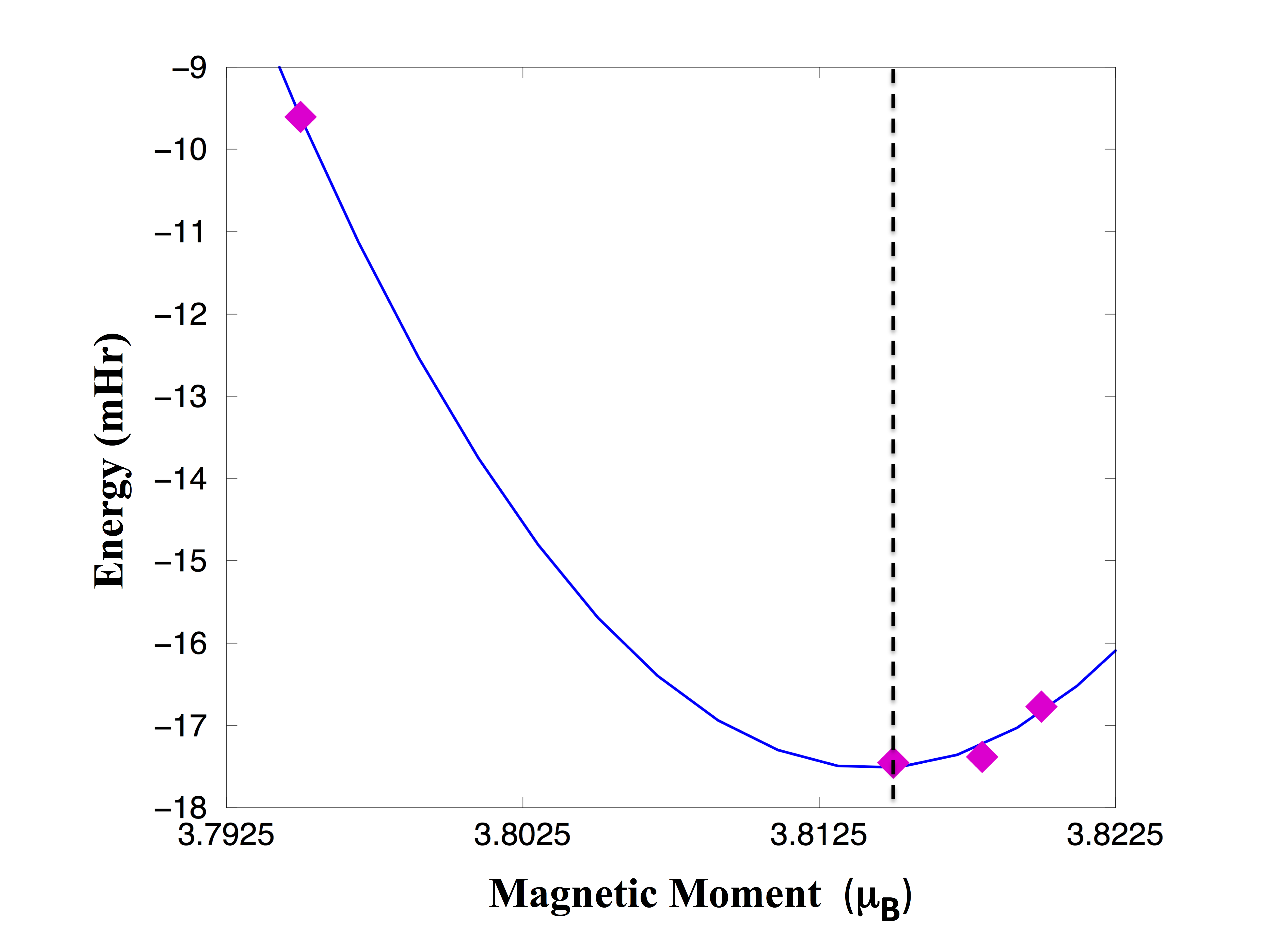}}
%\includegraphics[angle=0,width=0.5\columnwidth,clip]{xasfM5}}
%\centerline{\includegraphics[angle=0,width=1.0\columnwidth,clip]{dos-C}, }
\caption{
The total spin magnetic moment per unit cell vs the exchange
splitting $\Delta_{EX}$ dependence (A);
the total energy as a function the magnetic moment dependence,  $E^{tot}(M) = const + \alpha M^2 - \beta M^4$.
The total energy minimum position is marked by dashed line.}
\label{fig:2}
\end{figure} 

The exchange splitting $ \Delta_{\rm EX}$ in the Eq.~(\ref{eq:hamilt}) corresponds to the interorbital exchange energy 
 between the localized 4$f$ and itinerant $s$ and $d$ shells~\cite{Peters2014,piveta2020}. 
The $ \Delta_{\rm EX}$ can be estimated as 
  $$ \Delta_{\rm EX} =  2 J_{fs}S_{6s} + 2 J_{fd}S_{5d} \; , $$   
 where $J_{fs}$ and $J_{fd}$ are the interorbital exchange constants~\cite{piveta2020}. 
The spin-polarized DFT calculations with the magnetization directed along the $z$-axis yield
$\Delta_{\rm EX} \approx$ 10 meV, which can be taken as a
lower bound value for the interorbital exchange energy
~\cite{Peters2014}.

We performed the DFT+U(HIA) calculations 
treating $\Delta_{\rm EX}$  as a parameter in the Eq.~(\ref{eq:hamilt}).
In these spin-polarized calculations 
we applied  the DFT non-spin-polarized exchange-correlation potential to the $f$-states in the
Eq.~(\ref{eq:kohn_sham}), in order to exclude the contribution of  $f$-intraorbital exchange field
into the double-counting  $V_{dc}$. The spin-polarized functional is used for all other states.

We solve self-consistently the Eq.~(\ref{eq:kohn_sham}), and obtain dependence of the total spin magnetic moment per unit cell
$M(\Delta_{\rm EX})$
(see Fig.~\ref{fig:2}A) and the total energy $E^{tot}(\Delta_{\rm EX})$  Eq.~(\ref{totale})
on the magnitude of the $ \Delta_{\rm EX}$. Note that the upper bound limit
of $\Delta_{\rm EX} \approx$ 40 meV is set by reaching the saturation of the magnetic moment. 

The total energy vs the magnetic moment dependence $E^{tot}(M)$ is shown in Fig.~\ref{fig:2}B.
%Next, we make use of 
Using the Landau expansion~\cite{LL1980} of the magnetic energy,
$$E^{tot}(M) = const + \alpha M^2 - \beta M^4$$
we obtain the magnetic moment $M=\sqrt{\alpha \over {2 \beta}}$ 
which corresponds to the minimum of the $E^{tot}$. The corresponding value of $ \Delta_{\rm EX} \approx 20$ meV yields 
the value of the interorbital exchange energy in the Eq.~(\ref{eq:hamilt}). 
\begin{figure}[!htbp]
\includegraphics[angle=0,width=1.0\columnwidth,clip]{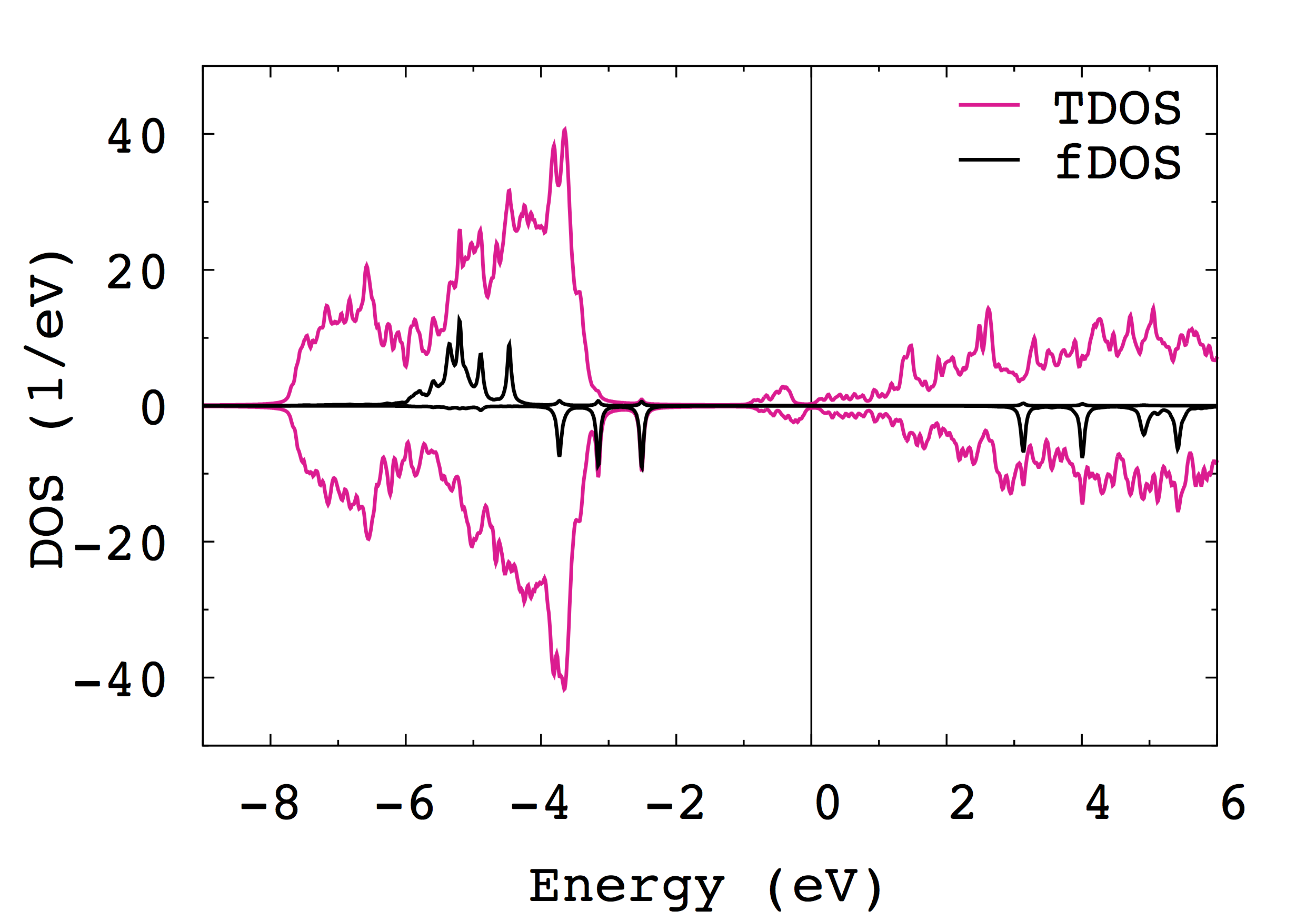}
\includegraphics[angle=0,width=1.0\columnwidth,clip]{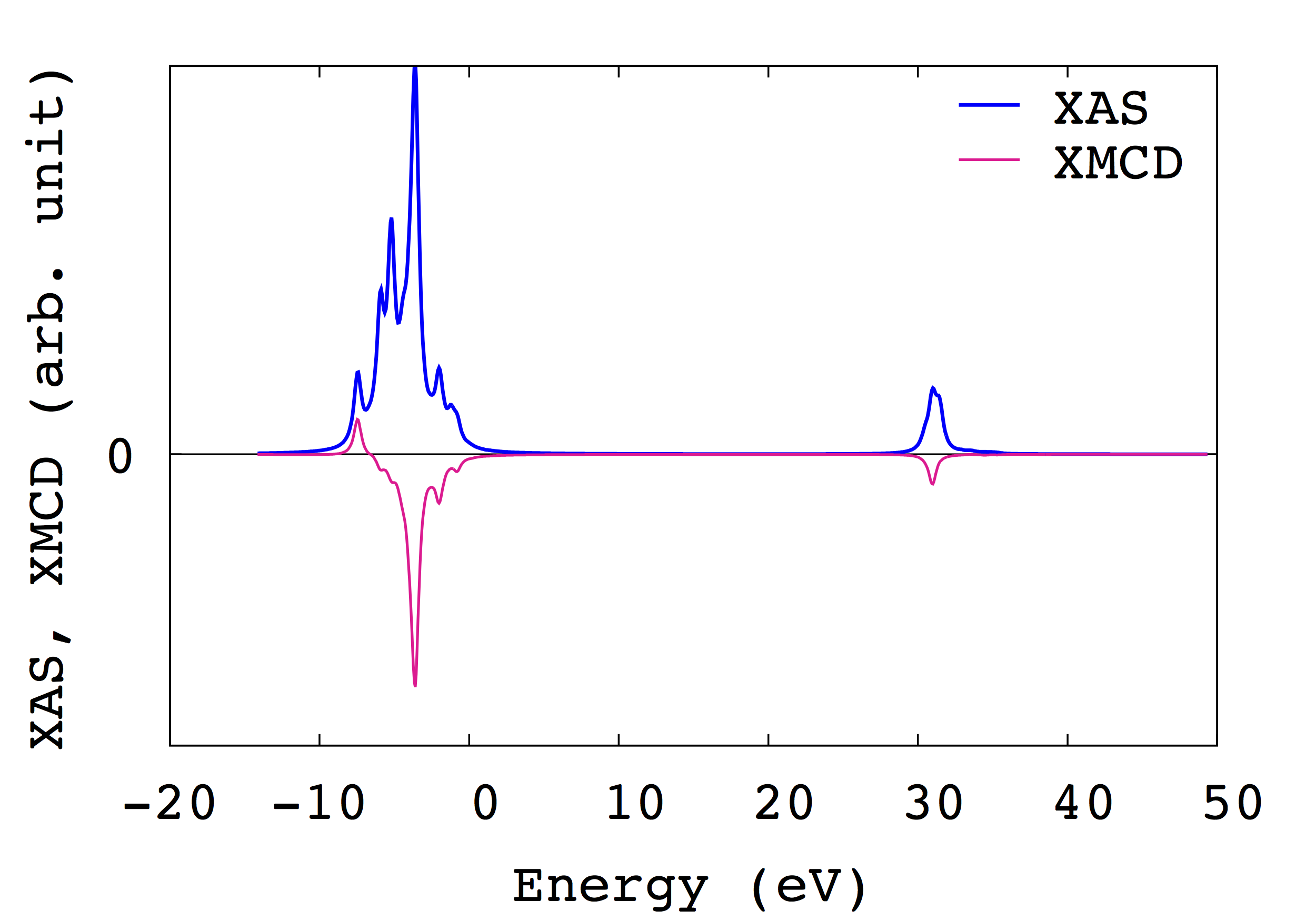}
\caption{
The spin-resolved total (TDOS) and the $f$-projected ($f$DOS) DOS (A);
%The spin-resolved spectral $f$DOS~(B);
the M-edge XAS and XMCD spectra (normal incidence) (B)
 for Dy@MgO(001)}
\label{fig:3}
\end{figure} 

 \begin{table}[!htbp] 
\caption{The $f$-electron occupation $n_f$ , spin $\langle M_S \rangle$, orbital $\langle M_L \rangle$,   $\langle M_S \rangle$ plus magnetic dipole $\langle M_D \rangle$  moments (in $\mu_B$), and the ratio $R_{LS} = {{\langle M_L \rangle} \over {\langle M_S \rangle + \langle M_D \rangle}}$ for the Dy adatom on MgO(001).
The nonzero Stevens parameters $B_k^q$ (in $\mu eV$).}
\begin{center}
\label{tab:1}
\begin{tabular}{ccccccc}
\hline
  &   $n_f$        &$\langle M_S \rangle$&$\langle M_L \rangle$&$\langle M_S \rangle$+$\langle M_D \rangle$&$R_{LS}$ \\
\hline
Dy@MgO           &   9.91      & 3.65                          & 5.92                            & 4.64                                                                 & 1.28       &  \\ 
%Dy@GR~\cite{shick2020}    &   9.88      & 3.74                          & 5.97                            & 4.69                                                                 & 1.27       &  \\ 
\hline
\end{tabular}
\\
\begin{tabular}{ccccccc}
\mbox{CEF} & $B_2^0$ & $B_4^0$ & $B_6^0$ & $B_4^4$ &$B_6^4$ \\
           & -20.55  & 0.23    & -0.02   &  1.81   & 0.04   \\
\hline
\end{tabular}
\end{center}
\end{table}
The calculated ground state $f$-electron occupation $n_f = {\rm Tr} [\hat{n}]$, magnetic  spin $ \langle M_S  \rangle = -2 \langle S_z \rangle \mu_B/\hbar = - {\rm Tr} [\hat{\sigma}_z \hat{n}] \mu_B/\hbar$,  orbital  $ \langle M_L  \rangle=-\langle L_z \rangle \mu_B/\hbar$, 
dipole $ \langle M_D  \rangle = -6 \langle T_z \rangle  \mu_B/\hbar$ moments, and  $R_{LS} = {{\langle M_L \rangle} \over {\langle M_S \rangle + \langle M_D \rangle}}$ value, the ratio of the orbital to the effective spin moment, are shown in Table~{\ref{tab:1}}. 
The itinerant part of the magnetization of 0.10 $\mu_B$
includes the Dy adatom  6$s$-states $m_{6s}$ = 0.02 $\mu_B$, and  5$d$-states  $m_{5d}$ = 0.02 $\mu_B$ magnetic moments. Note that the calculation of these moments is associated with some uncertainty, and  depends on the choice of the Dy adatom muffin-tin radius.  

The total (TDOS) and $f$-projected ($f$DOS) DOS  calculated from the solutions of the  Eq.(\ref{eq:kohn_sham}) are shown in  Fig.~\ref{fig:3}~(A).
The MgO band gap is at $\approx$ 3-to-1 eV below the Fermi level. The sharp 
4$f$-spin-$\downarrow$ peaks are located at the top of MgO valence band gap.
The smooth TDOS peak  $\approx$ 1 eV below the Fermi level   has a capacity of 2 electrons which are transfered from the Dy adatom to the MgO substrate.
%The spin-resolved spectral $f$DOS calculated making use of Eq.~(\ref{eq:gf}) are shown in %Fig.~\ref{fig:2}~(B).

The $f$-electron occupation $n_f=9.91$ is consistent with  the $f^{10}$ configuration obtained from Eq.~(\ref{eq:hamilt}), and defines 
the Dy adatom valence as  Dy$^{2+}$. We used the Eq.~(\ref{eq:hamilt}), with the self-consistently determined parameters  as an input for the  Quanty code~\cite{quanty} to estimate the M-edge XAS and XMCD spectra (see for details Supplemental material).  
The computed spectra (Fig.~\ref{fig:3}B) are in a reasonable agreement with available experimental data~\cite{Donati2021}.

\begin{figure}[!hb]
%\centerline{\resizebox*{12cm}{!}{\includegraphics[angle=0]{fig3.png}}}
\centerline{\includegraphics[angle=0,width=1.0\columnwidth,clip]{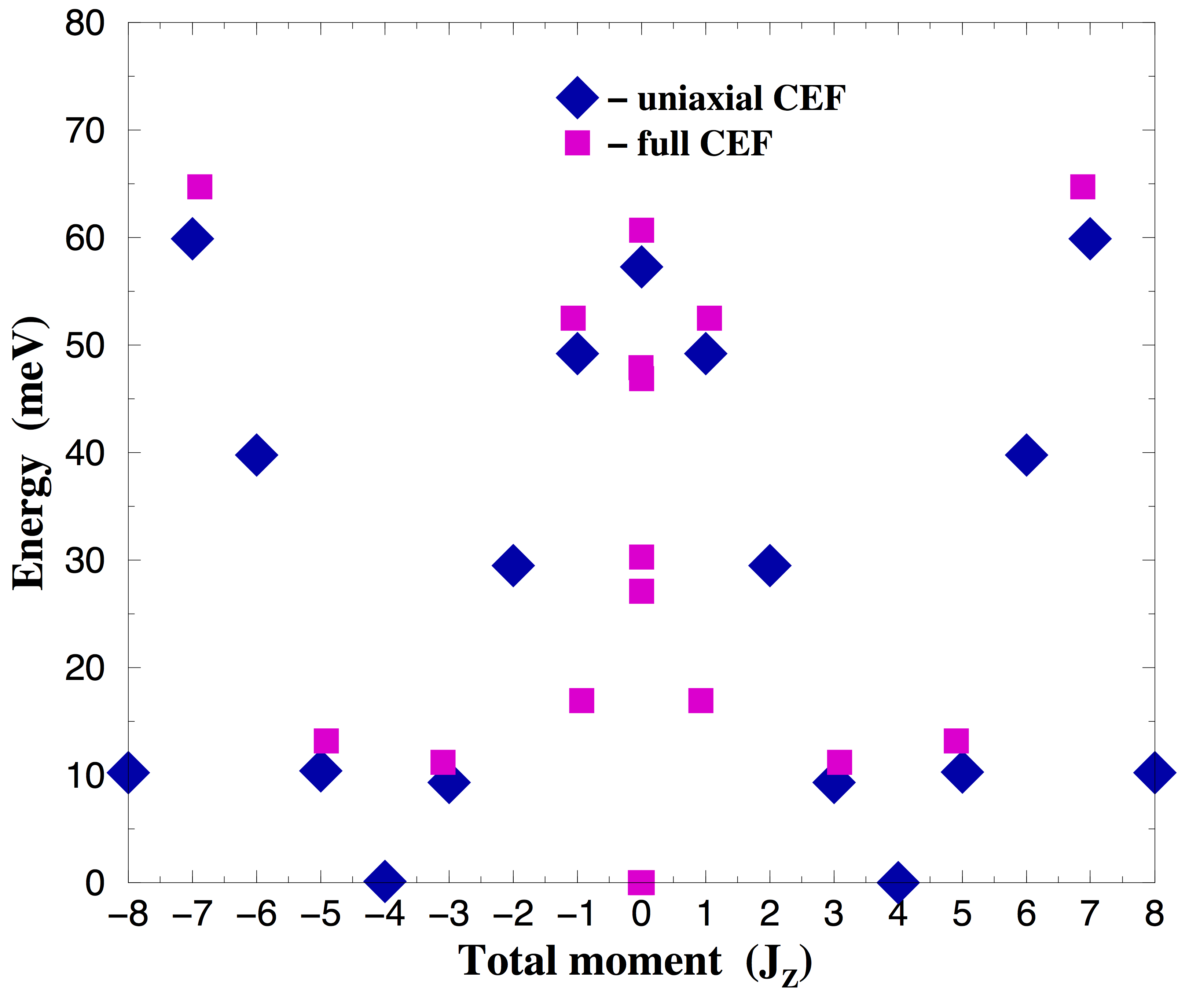}}
\caption{Scheme of quantum many-body levels of the lowest $J=8.0$ multiplet obtained from the 
solutions  of  Eq.~~(\ref{eq:hamilt}) ($\Delta_{\rm ex}=0$)
with the $\Delta_{\rm CF}$ parameters taken from spin-polarized calculations (squares); energy diagram 
of the CF Hamiltonian with the uniaxial CEF parameters only (diamonds).}
\label{fig:4}
\end{figure} 

The scheme of quantum many-body levels of the lowest $J=8.0$ multiplet obtained from the 
solutions of Eq.~(\ref{eq:hamilt}) is shown in  Fig.~\ref{fig:4}. 
Without an external magnetic field, the lowest energy state of Eq.~(\ref{eq:hamilt}) is 
a singlet $\ket{J=8.0, J_z= 0.0 }$ state. There is another 
$\ket{J=8.0, J_z = 0.0 }$ singlet with the energy of 0.06 meV above the ground state. 
Leaving the only uniaxial (diagonal)
contributions to the $\Delta_{\rm CF}$ yields the $\ket{J=8.0, J_z=  \pm 4.0 }$ ground state
(cf. Fig.~\ref{fig:4}).

The $\Delta_{\rm CF}$ matrix calculated in the DFT+U(HIA)
is used to build the CF hamiltonian~\cite{shick2019} for the Dy@MgO(001),
\begin{equation}
\hat{H}_{CF}=\sum_{kq} B_k^q \hat{O}_k^q \; ,
\label{eq:CEF}
\end{equation}
where $\hat{O}_k^q$ are the Stevens operator equivalents,  and $B_k^q$,  the Stevens crystal field parameters (in standard notations) for given $k$ and $q$. The five evaluated non-zero Stevens parameters,
$B_2^0$, $B_4^0$, $B_6^0$, $B_4^4$, and $B_6^4$
are shown in Table.~\ref{tab:1}. The energy diagrams 
of the CF hamiltonian~(\ref{eq:CEF}) are shown in Fig.~\ref{fig:S1}  (see Supplemental material.  Both diagrams, with the full set of the CF parameters,
and with the first three uniaxial CF parameters are shown). 
It is seen that the CF solutions approximate reasonably well 
the many-body solutions of the Eq.~(\ref{eq:hamilt}) shown in Fig~\ref{fig:4}.

The first three parameters $B_2^0$, $B_4^0$, $B_6^0$ yield the uniaxial
splitting between different $J_z$ eigenstates in Eq.~(\ref{eq:CEF}) with the $\ket{J=8.0, J_z=  \pm 4.0 }$ ground state, and correspond to diagonal
contributions to the $\Delta_{\rm CF}$.  The energy difference between the lowest and highest  $J_z$ levels, the so-called zero field splitting (ZFS) of 65 meV is found, 
which is related to the uniaxial magnetic anisotropy~\cite{baltic2018}.
The transverse $B_4^4 O_4^4$ term in the CF hamiltonian connects  
the $\ket{J=8.0, J_z=  \pm 4.0 }$ states so that the quantum tunneling of the magnetization (QTM) occurs between these two states, and the resulting
$\ket{J=8.0, J_z = 0 }$ ground state corresponds to the ``in-plane" magnetic moment orientation. 
It explains an absence of the remanent magnetization in Dy@MgO(001) observed experimentally~\cite{Donati2021}.

%{\em Summary and outlook}
 To conclude, the electronic structure and magnetism of  individual Dy atom adsorbed on the MgO(001) substrate is
 investigated using the combination of the density functional theory with the Hubbard-I approximation to the Anderson impurity model.
 The divalent  Dy$^{2+}$  adatom is found with a singlet $\ket{J=8.0, J_z= 0.0 }$ ground state. 
 The calculated XAS and XMCD spectra are in reasonable agreement with available experimental data.
 No remanent magnetization is found due to QTM, in agreement with experimentally observed butterfly-type magnetic hysteresis loop.
 %Our studies can provide a viable route for further investigation and prediction of the rare-earth single atom  magnets.

%\section{Acknowledgments}
We acknowledge stimulating discussions with J. Kolorenc and A. Yu. Denisov. 
Financial support was provided by Operational Programme Research, Development and Education financed by European Structural and Investment Funds and the Czech Ministry of Education, Youth and Sports 
(Project No. SOLID21 - CZ.02.1.01/0.0/0.0/16$_{-}$019/0000760), by the Czech Science Foundation (GACR) Grant
No. 22-22322S, and from the Israeli Ministry of Aliyah and Integration Grant Ref.:140636.
    
\bibliographystyle{iopart-num}
%\bibliography{refs}

\clearpage
\appendix
\section{Supplemental Material}
\subsection{Computational details}
In the DFT+U(HIA) FP-LAPW calculations, 49 special k-points in the two-dimensional  Brillouin zone
were used, with Gaussian smearing for k-points weighting.
The ``muffin-tin'' radii of $R_{MT} \; = \; 2.70 \; a.u.$ for Dy,
$R_{MT}=1.20$ a.u. for O,  $R_{MT}=2.00$ a.u. for Mg  were used.
The LAPW basis cut-off is defined by the condition
$R^{Dy}_{MT} \times K_{max} \; = \; 8.10$
(where $K_{max}$ is the cut-off for LAPW basis set). 

The CF matrix $\Delta_{\rm CF}$ in Eq.(\ref{eq:hamilt}) is obtained by 
projecting the self-consistent solutions of Eq.(\ref{eq:kohn_sham})  into
the $\{\phi_{\gamma}\}$ local $f$-shell basis, giving the ``local Hamiltonian"
\begin{eqnarray}
[H_{loc}]_{\gamma\gamma'} &=&
\int_{\epsilon_b}^{\epsilon_t} {\rm d} \epsilon \, \epsilon 
             [N(\epsilon)]_{\gamma \gamma'}   \nonumber \\ 
    &\approx& \epsilon_0 \delta_{\gamma \gamma'}
+ [\xi {\bf l}\cdot{\bf s} + \Delta_{\rm CF}]_{\gamma \gamma'}  + [{V_{U}}]_{\gamma \gamma'} \, ,
 \label{eq:hloc}
\end{eqnarray}
where $[N(\epsilon)]_{\gamma_1 \gamma_2}$ is the $f$-projected 
density of states (fDOS) matrix 
$$[N(\epsilon)]_{\gamma_1 \gamma_2} \; = \; - \pi^{-1} {\rm Im} {[G(z)_{\rm DFT+U}]}_{\gamma_1 \gamma_2} \; ,$$
$\epsilon_b$ is the bottom of the valence band, $\epsilon_t$ is the upper cut-off,
which is naturally defined by the condition
 $\int_{\epsilon_b}^{\epsilon_t} {\rm d} \epsilon \, {\rm Tr} {[N(\epsilon)]} \; = 14$,
and $\epsilon_0$ is the mean position of the non-interacting $5f$
level. 
The matrix  $\Delta_{\rm CF}$ is then obtained by removing the interacting DFT+$U$
potential  $[{V_{U}}]_{\gamma \gamma'}$ and SOC  $[\xi {\bf l}\cdot{\bf s}]_{\gamma \gamma'}$ from 
$H_{loc}$ Eq.( \ref{eq:hloc}).

\subsection{Calculation of XAS and XMCD spectra}
We used the ionic hamiltonian, Eq.~(\ref{eq:hamilt}), with the self-consistently determined $\Delta_{CF}$ parameters  as an input for the  Quanty code~\cite{quanty} to estimate the M-edge XAS and XMCD spectra. 
In these calculations, the exchange field $\Delta_{\rm EX}$ is replaced with the external magnetic field $B_z = 6.8$~T typical in the 
experimental XMCD measurements~\cite{singha2017}. The 3d--4f Coulomb interaction is parametrized with Slater integrals computed with the Cowan's Hartree--Fock code~\cite{cowan} and then reduced to 80\% to approximately account for screening (Table~\ref{tab:3d4fSlater}). The 3d spin-orbit coupling $\xi_{3d}=14.4$~eV is taken from the same Hartree--Fock calculations.

\begin{table}[H]
\caption{\label{tab:3d4fSlater}Slater integrals defining the 3d--4f Coulomb interaction as computed with the Cowan's code~\cite{cowan} for the XAS final state 3d\textsuperscript{9}4f\textsuperscript{11}. The Hartree--Fock values are reduced to 80\% to account for screening. All values are shown in eV.}
\renewcommand{\arraystretch}{1.3}
\begin{ruledtabular}
\begin{tabular}{lccccc}
 & $F_2$ & $F_4$ & $G_1$ & $G_3$ & $G_5$ \\
\hline
%3d\textsuperscript{9}4f\textsuperscript{10} &
% 7.79 & 3.66 & 5.64 & 3.31 & 2.29 \\
3d\textsuperscript{9}4f\textsuperscript{11} &
 7.36 & 3.44 & 5.28 & 3.10 & 2.14 \\
\end{tabular}
\end{ruledtabular}
\end{table}
%\clearpage
\subsection{Crystal-field model parameters}
\begin{figure}[H]
%\centerline{\resizebox*{12cm}{!}{\includegraphics[angle=0]{fig3.png}}}
\centerline{\includegraphics[angle=0,width=1.0\columnwidth,clip]{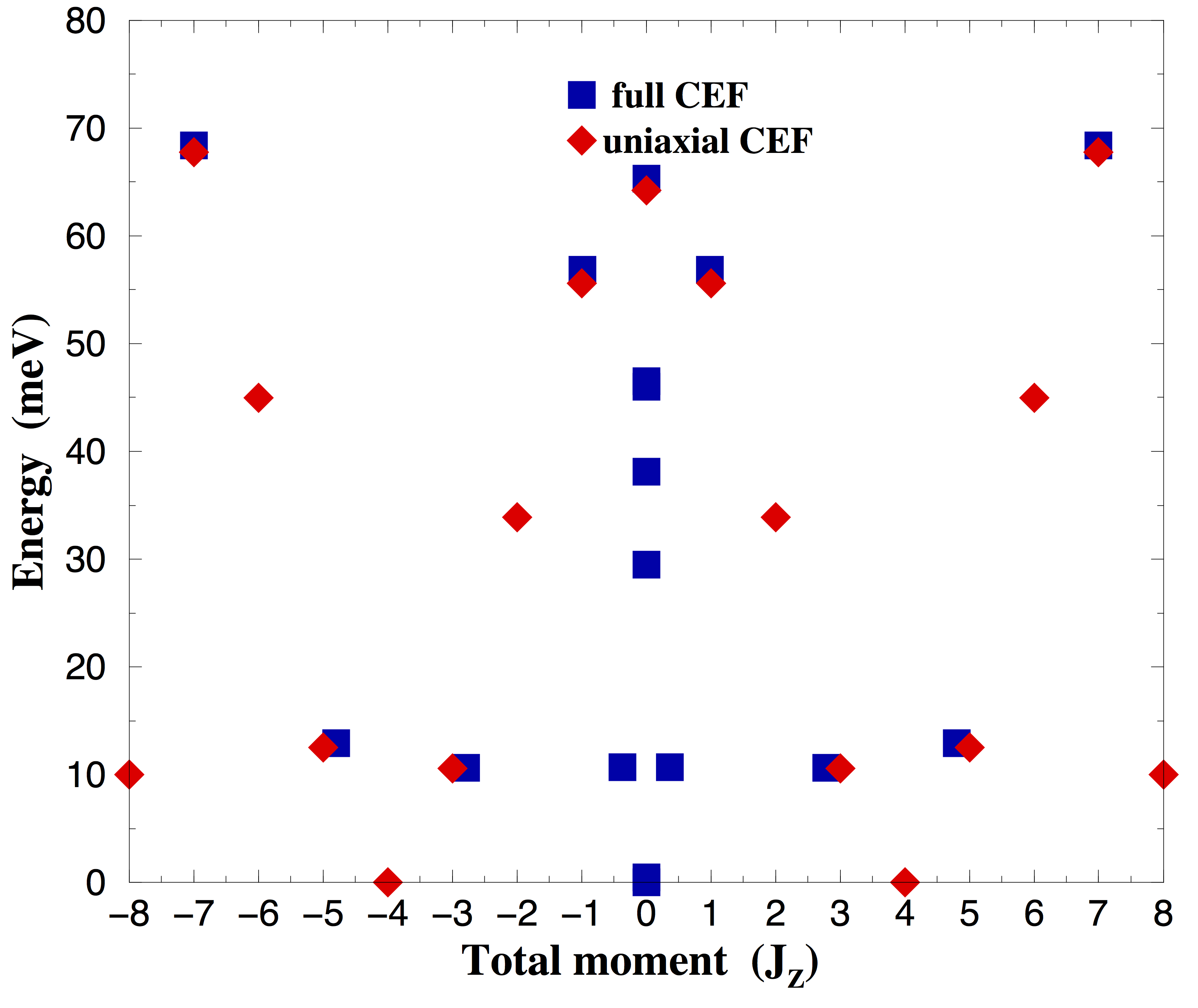}}
\caption{The energy diagram 
of the CF Hamiltonian with the CF parameters from Table~I.  (squares), and the uniaxial CEF only (diamonds).}
\label{fig:S1}
\end{figure}

The energy diagrams 
of the CF Hamiltonian~(\ref{eq:CEF}) with the CF parameters from Table~I of the main text.
are shown in in Fig.~\ref{fig:S1}. Both diagrams, with the full set of five Stevens parameters,
and with the first three uniaxial CF parameters are shown. 
It is seen that the CF solutions approximate reasonably well 
the many-body solutions of the Eq.~(\ref{eq:hamilt}) shown in Fig. 4 of the main text. 
The first three CF parameters $B_2^0$, $B_4^0$, $B_6^0$ yield the uniaxial splitting
of different $J_z$ eigenstates in Eq.~(\ref{eq:CEF}), and  the $\ket{J=8.0, J_z=  \pm 4.0 }$ ground state. The energy difference between the lowest and highest  $J_z$ levels (ZFS) of 65 meV is found.
Once the nonzero transverse CF parameters $B_4^4$, and $B_6^4$ are included in Eq.(\ref{eq:CEF}), 
the ground state becomes a singlet $\ket{J=8.0, J_z= 0 }$ with another singlet
$\ket{J=8.0, J_z = 0 }$ with the energy of 0.5 meV above the ground state. 

\end{document}